\documentclass[sigconf, nonacm]{acmart}

\usepackage[symbol]{footmisc}

\AtBeginDocument{%
  \providecommand\BibTeX{{%
    \normalfont B\kern-0.5em{\scshape i\kern-0.25em b}\kern-0.8em\TeX}}}

\begin{document}

\title{Enhancing Cross-Sectional Currency Strategies by Context-Aware Learning to Rank with Self-Attention}



\author{Daniel Poh}
\affiliation{%
  \institution{Department of Engineering Science, Oxford-Man Institute of Quantitative Finance, University of Oxford}
  \streetaddress{Eagle House, Walton Well Road, OX2 6ED}
  \city{Oxford}
  \country{United Kingdom}}
\email{dp@robots.ox.ac.uk}

\author{Bryan Lim}
\affiliation{%
  \institution{Department of Engineering Science, Oxford-Man Institute of Quantitative Finance, University of Oxford}
  \streetaddress{Eagle House, Walton Well Road, OX2 6ED}
  \city{Oxford}
  \country{United Kingdom}}
\email{blim@eng.oxon.org}

\author{Stefan Zohren}
\affiliation{%
  \institution{Department of Engineering Science, Oxford-Man Institute of Quantitative Finance, University of Oxford}
  \streetaddress{Eagle House, Walton Well Road, OX2 6ED}
  \city{Oxford}
  \country{United Kingdom}}
\email{stefan.zohren@eng.ox.ac.uk}

\author{Stephen Roberts}
\affiliation{%
  \institution{Department of Engineering Science, Oxford-Man Institute of Quantitative Finance, University of Oxford}
  \streetaddress{Eagle House, Walton Well Road, OX2 6ED}
  \city{Oxford}
  \country{United Kingdom}}
\email{sjrob@robots.ox.ac.uk}

\renewcommand{\shortauthors}{Poh D., Lim B., Zohren S. and Roberts S.}

\begin{abstract}

The performance of a cross-sectional currency strategy depends crucially on accurately ranking instruments prior to portfolio construction. 
While this ranking step is traditionally performed using heuristics, or by sorting the outputs produced by pointwise regression or classification techniques, strategies using Learning to Rank algorithms have recently presented themselves as competitive and viable alternatives. 
Although the rankers at the core of these strategies are learned globally and improve ranking accuracy on average, they ignore the differences between the distributions of asset features over the times when the portfolio is rebalanced. 
This flaw renders them susceptible to producing sub-optimal rankings, possibly at important periods when accuracy is actually needed the most. For example, this might happen during critical risk-off episodes, which consequently exposes the portfolio to substantial, unwanted drawdowns. We tackle this shortcoming with an analogous idea from information retrieval: that a query's top retrieved documents or the \textit{local ranking context} provide vital information about the query's own characteristics, which can then be used to refine the initial ranked list. In this work, we use a context-aware Learning-to-rank model that is based on the Transformer architecture to encode top/bottom ranked assets, learn the context and exploit this information to re-rank the initial results.
Backtesting on a slate of 31 currencies, our proposed methodology increases the Sharpe ratio by around 30\% and significantly enhances various performance metrics.
Additionally, this approach also improves the Sharpe ratio when separately conditioning on normal and risk-off market states.

\end{abstract}

\keywords{momentum strategies, learning to rank, neural networks, machine learning, information retrieval, foreign exchange markets, portfolio construction}

\maketitle

\section{Introduction}
Cross-sectional strategies are a popular trading style, with numerous works in academic finance documenting technical variations and their use across different asset classes \cite{bazDissectingInvestmentStrategies2015}. Unlike time-series methods which adopt a narrow, longitudinal focus and trade individual assets independently \cite{moskowitzTimeSeriesMomentum2012}, cross-sectional strategies cover a broader slate of instruments and typically involve buying assets with the highest expected returns (Winners) while simultaneously selling those with the lowest (Losers). 
The classical cross-sectional momentum (CSM) of \cite{jegadeeshReturnsBuyingWinners1993} for instance, was originally applied to equities and assumes the persistence of returns -- trading some top segment of stocks against an equally-sized bottom segment after ranking them by returns over the past 12-months. Here we focus on the foreign exchange (FX) market which is more liquid and enjoys larger trading volumes as well as lower transaction costs \cite{menkhoffCurrencyMomentumStrategies2012}. The confluence of these factors, and the fact that participants are typically sophisticated investors, raises the bar for generating consistent excess returns over time \cite{menkhoffCurrencyMomentumStrategies2012}.

Along with the proliferation of machine learning, numerous cross-sectional strategies incorporating advanced prediction techniques such as \cite{kimEnhancingMomentumStrategy2019, guEmpiricalAssetPricing2018, guAutoencoderAssetPricing2019} have been developed. One notable subset of these models  employ Learning to Rank (LTR) algorithms. This class of methods explicitly account for the crucial \textit{expected order} of returns, which is in turn an important driver of cross-sectional trading performance. The usefulness of this approach is empirically validated by \cite{pohBuildingCrossSectionalSystematic2021} who demonstrate the superiority of CSM techniques employing LTR over conventional baselines on US equities.

LTR is an active area of research within Information Retrieval (IR), and focuses on training models using supervised techniques to perform ranking tasks \cite{liLearningRankInformation2014}. Among the numerous studies seeking to develop better LTR algorithms, one particular broad branch with works such as \cite{aiLearningGroupwiseMultivariate2019, indrakantiInfluenceNeighborhoodPreference2019, sohnLearningStructuredOutput2015, pobrotynContextAwareLearningRank2020} concentrate on incorporating inter-item dependencies not just at the loss level, but also within the scoring function. Within this area, some studies \cite{belloSeq2SlateRerankingSlate2018, taozhuangGlobalRerank2018, peiPersonalizedRerankingRecommendation2019, aiLearningDeepListwise2018} use \textit{re-ranking}, where a sorted list from a previous ranking algorithm is used as input to model the complex interactions among items. From this, a scoring function is learned and is used to refine the original sorted list. 
Adopting this approach, \cite{aiLearningDeepListwise2018} tackle a weakness of globally learned ranking functions -- that is, they can sort objects accurately on average, but fare poorly over certain queries as they ignore query-specific feature distributions. Motivated by the idea that a query's top retrieved documents or the \textit{local ranking context} contains information that can boost the performance of LTR systems, the authors develop the DLCM (Deep Listwise Context Model). 
Central to the DLCM (which was a state-of-the-art method for encoding feature vectors \cite{peiPersonalizedRerankingRecommendation2019}) is a recurrent neural network (RNN) that is used to encode the local ranking context obtained by an initial sort for re-ranking.
Recognising the limitations of the RNN architecture used by the DLCM, \cite{pobrotynContextAwareLearningRank2020} replace it with a Transformer encoder \cite{vaswaniAttentionAllYou2017}. By incorporating positional embeddings as a further modification, they show that this approach is suitable for re-ranking and report significant improvements over DLCM.

\begin{figure}[t!]
  \centering
  \includegraphics[width=\linewidth]{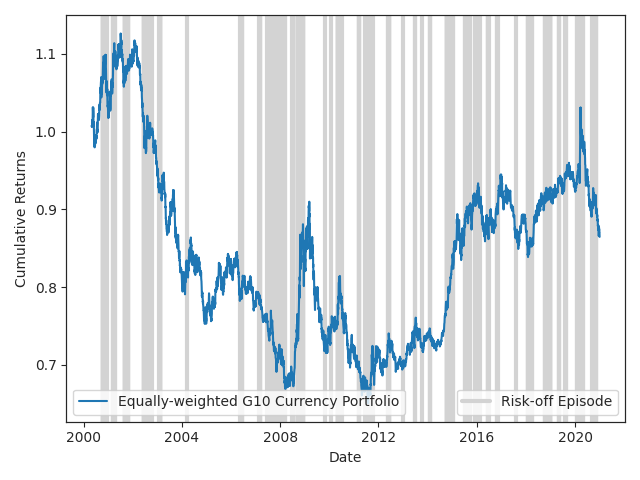}
  \caption{Risk-off episodes (e.g. the prolonged period of risk-aversion around the GFC in 2008) appear to coincide with large moves in the cumulative returns of an equally-weighted portfolio of G10 currencies. Risk-off states have also increased in frequency since then.}\label{fig:curr_riskoff}
\end{figure}
By the same token, current CSM strategies built on top of globally trained LTR algorithms are able to sort assets accurately on average. However, relying on these global rankers subjects the broader strategy to inaccurate rankings at certain points in time -- potentially at instances when accuracy is actually needed the most, such as during critical (and increasingly frequent) risk-off episodes as shown in Figure \ref{fig:curr_riskoff}\footnote{We define a single risk off episode or state as an instance when the VIX is 5\% points higher than its rolling 60-day moving average; Normal states are then simply states which are not risk-off states. All currencies are expressed versus the USD, and the vertical lines indicating risk-off states have been broadened to improve visibility.}.
Following \cite{pobrotynContextAwareLearningRank2020}, we use a context-aware LTR model that is based on the Transformer which has established state-of-the-art performance on the  MSLR-WEB30K\footnote{This is large scale LTR data set released by Microsoft and is often used to benchmark the performance of ranking algorithms.} data set. Starting with a sorted list obtained from a previous ranking algorithm, we cast our problem in a similar supervised setting utilised by LTR and re-rank this list using our proposed model. These results are then concretely evaluated against a mix of LTR and conventional benchmarks. On a set of 31 currency pairs, we demonstrate how the performance of the original strategy can be enhanced with our approach.

\section{Related Works}

\subsection{Cross-sectional Currency Momentum}
Momentum strategies can be classified as being either time-series or cross-sectional. In time-series momentum, which was first proposed by \cite{moskowitzTimeSeriesMomentum2012}, an instrument's trading rule relies only on previous returns. 
With cross-sectional momentum (CSM), the relative performance of assets is the essence of the strategy -- after ranking securities based on some scoring model, the long/short portfolios are constructed by trading the extremes (e.g. top and bottom deciles) of the ranked list. Since \cite{jegadeeshReturnsBuyingWinners1993}, which demonstrate the profitability of this strategy on US equities, the literature has been galvanised by a broad spectrum of works  -- ranging from technical refinements spanning varying levels of sophistication \cite{bazDissectingInvestmentStrategies2015, kimEnhancingMomentumStrategy2019, pirrongMomentumFuturesMarkets2005, guAutoencoderAssetPricing2019} to reports documenting the ubiquity of momentum in different asset classes and markets \cite{rouwenhorstInternationalMomentumStrategies1998, griffinMomentumInvestingBusiness2003, chuiIndividualismMomentumWorld2010, grootCrosssectionStockReturns2012, erbStrategicTacticalValue2006, lebaronTechnicalTradingRule1996}. 
Unlike the extensive works focused on equity momentum, currency momentum is centred on the time-series of individual currency pairs and is often cast as "technical trading rules" for which \cite{menkhoffObstinatePassionForeign2007} provides a broad overview. While studies such as \cite{burnsideCarryTradeMomentum2011, okunevMomentumBasedStrategiesStill2003} offer evidence of CSM being profitable in FX markets, their results involve trading a narrow cross-section involving only major pairs and lack a unifying analysis that explains their returns. 
\cite{menkhoffCurrencyMomentumStrategies2012} addresses this research gap and also demonstrate that portfolios constructed by trading the winner/loser segments (analogous to \cite{jegadeeshReturnsBuyingWinners1993} in the equity literature) can generate high unconditional excess returns. In \cite{bazDissectingInvestmentStrategies2015}, the authors adopt a more sophisticated approach by using volatility-scaled moving-average convergence divergence (MACD) indicators as inputs. 
Beyond this and unlike the comparatively abundant work on equity momentum, we find little published in terms of constructing currency CSM portfolios.

\subsection{Learning to Rank}
Learning to Rank (LTR) is an active research area within information retrieval focusing on developing models that learn how to sort lists of objects in order to maximise utility. Its prominence grew in parallel with the accessibility of modern computing hard- and software, and its algorithms utilise sophisticated architectures that allow it to learn in a data-driven matter -- unlike their predecessors such as BM25 \cite{robertsonSimpleEffectiveApproximations1994} and LMIR \cite{ponteLanguageModelingApproach1998} which require no training but which instead need handcrafting and explicit design \cite{liLearningRankInformation2014}. We point the interested reader to \cite{liuLearningRankInformation2011} for a comprehensive overview.

Given the widespread adoption of LTR across numerous applications such as search engines, e-commerce \cite{santuApplicationLearningRank2017} and entertainment \cite{pereiraOnlineLearningRank2019}, there is motivation both in academia and industry to develop better models.  
Commonly used methods typically learn a ranking function that assigns scores to items individually (i.e. without the \textit{context} of other objects in the list). Although these models are trained by optimising a loss function which may capture the interactions among items, they score them individually at inference-time without accounting for any mutual influences among the items \cite{pobrotynContextAwareLearningRank2020}. 
To this end, a number of studies \cite{aiLearningGroupwiseMultivariate2019, indrakantiInfluenceNeighborhoodPreference2019, sohnLearningStructuredOutput2015, pobrotynContextAwareLearningRank2020} overcome this disadvantage by modelling inter-item dependencies at both the loss level and within the scoring function. Some works \cite{peiPersonalizedRerankingRecommendation2019, aiLearningDeepListwise2018, taozhuangGlobalRerank2018, yinRankingRelevanceYahoo2016} within this area achieve this with re-ranking. The main idea is to explicitly model the interactions between sorted items provided by some previous ranking algorithm, and subsequently use this information to refine the same sorted list.
In \cite{aiLearningDeepListwise2018}, re-ranking with the top retrieved items or local ranking context is used to remedy the deficiency of globally-learned ranking functions ignoring differences between feature distributions across queries. Specifically, the authors note that depending on query characteristics, pertinent items for different queries are often distributed differently in the feature space. Hence, while a global ranker is able to sort accurately on average, it may struggle to produce optimal results for certain queries. Inspired by \cite{LavrenkoCroftRelevanceModels2001, ZhaiLaffertyPseudoRelModels2001, robertsonRelevanceWeightingSearch1976} documenting the effectiveness of the local ranking context in boosting the performance of text-based retrieval systems, \cite{aiLearningDeepListwise2018} propose the DLCM model. The DLCM uses an RNN to sequentially encode the features of top results in order to learn a local context model, which is later applied to re-rank the same top results.
A major drawback with RNN-based approaches however, is that the encoded feature information degrades with encoding distance. Motivated by Transformer \cite{vaswaniAttentionAllYou2017} architectures as a workaround, \cite{peiPersonalizedRerankingRecommendation2019} and \cite{ pobrotynContextAwareLearningRank2020} utilise an adapted variant -- exploiting both the self-attention mechanism that learns the inter-item dependencies without any decay over distance, as well as the encoding procedure that allows for parallelization \cite{peiPersonalizedRerankingRecommendation2019}. By incorporating positional encodings, \cite{pobrotynContextAwareLearningRank2020} demonstrates that this modification enables the model to perform re-ranking.

Surveying the LTR-related literature on finance, we note that existing works such as \cite{pohBuildingCrossSectionalSystematic2021, songStockPortfolioSelection2017, wangStockRankingMarket2018} are essentially based on globally learned rankers and are therefore affected by the concerns discussed earlier.
Given that context-aware models based on Transformers have been successfully used for re-ranking \cite{pobrotynContextAwareLearningRank2020, peiPersonalizedRerankingRecommendation2019}, we adopt a similar approach 
of applying a context-aware ranker to refine the initial results of popular LTR algorithms such as ListNet \cite{caoLearningRankPairwise2007} and LambdaMART \cite{burgesRankNetLambdaRankLambdaMART2010} and show that this improves performance.

\section{Problem Definition}

For a given portfolio of currencies that is rebalanced daily, the returns for a cross-sectional momentum (CSM) strategy at day $t$ can be expressed as follows:
\begin{align}
r^{CSM}_{t, t+1} = \frac{1}{n} \sum^{n}_{i=1} S^{(i)}_t \left ( \frac{\sigma_{tgt}}{\sigma^{(i)}_t} \right ) r^{(i)}_{t, t+1},
\label{eqn:csm_rets} 
\end{align}
where $r^{CSM}_{t,t+1}$ denotes the realised portfolio returns going from day $t$ to $t+1$, $n$ refers to the number of currency pairs in the portfolio and $S^{(i)}_t \in \{-1, 0, 1\}$ signifies the cross-sectional momentum signal or trading rule for pair $i$.
The annualised target volatility $\sigma_{tgt}$ is set at 15\% and asset returns are scaled with $\sigma^{(i)}_t$ which is an estimator for ex-ante daily volatility. For simplicity we use a rolling exponentially weighted standard deviation with a 63-day span on daily returns for $\sigma^{(i)}_t$, and note that other sophisticated methods such as GARCH \cite{BollerslevGARCH} can be employed.

\begin{figure*}[t!]
  \centering
  \includegraphics[width=\linewidth]{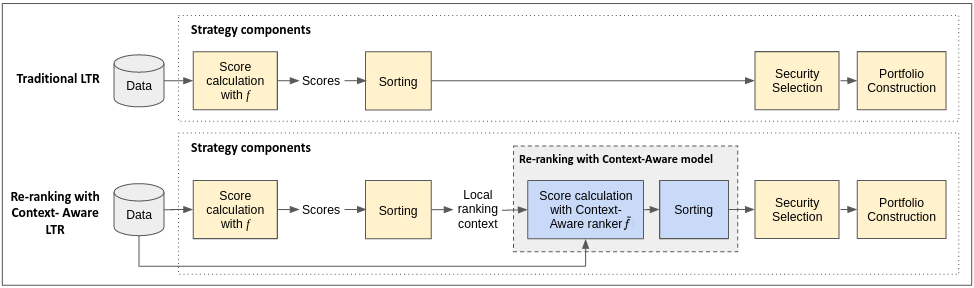}
  \caption{Flow diagrams of both traditional LTR (upper half) and the context-aware model (lower half) applied to the CSM strategy. The latter exploits the ranking context provided by some previous ranking algorithm to re-rank the initial set of scores provided by the same algorithm.}\label{fig:fw_compare}
\end{figure*}

\subsection{Problem Formulation} \label{sec:csm_framework} 

We now formalise the problem setup of adapting the current LTR framework with information provided by top retrieved items. The CSM strategy can be framed similar to a LTR problem\footnote{For a background on the parallels between the LTR frameworks as applied in the IR and CSM settings, we refer the reader to \cite{pohBuildingCrossSectionalSystematic2021}.}, where assets are sorted based on their momentum signals to maximise trading profits when the highest/lowest ranked instruments are traded. Given a specific time $t$ at which the portfolio is rebalanced, the feature vector of an instrument $c$ that is available for trading can be represented as $\textbf{x}_{(t, c)}$. Conventional LTR algorithms assume the existence of a global scoring function $f$ that, once learned, takes $\textbf{x}_{(t, c)}$ as input and then generates a ranking score for this object. For the same function parameterised by $\theta$, learning its optimal form is achieved by minimising the empirical loss:
\begin{equation}
    \mathcal{L}(\theta) = \sum_{t\in\tau} \ell \Bigl ( 
    \bigl \{ y_{(t, c)}, f(\textbf{x}_{(t, c)}; \theta) \bigr | c \in \mathcal{C} \bigr \}
    \Bigr )  \label{eqn:ltr_emp_loss}
\end{equation}
where $\tau$ represents all the times during which the portfolio is rebalanced, $\ell$ is the loss function computed with the ground truth $y_{(t, c)}$ (the sorted decile labels for the \textit{next} period) and the predicted score  $f(\textbf{x}_{(t, c)}; \theta)$. Lastly, $\mathcal{C}$ denotes the universe of tradeable currencies.

Suppose we now have $I(R^m_t, X^m_t, f(\cdot; \theta))$\footnote{\cite{aiLearningDeepListwise2018} represents the local ranking context with $I(R^m_t, X^m_t)$, but we include the dependence on $f(\cdot;\theta)$ as different ranking models produces different top/bottom retrieved results, and to also better distinguish $f$ from the proposed  context-aware ranking model  $\tilde{f}(\cdot;\tilde{\theta})$.}  which is the local ranking context provided by some previous ranking model $f$. 
Here, $R^m_t = \{ \textrm{top-}\textit{m } c \textrm{ sorted by } f({\textbf{x}}_{(t, c)}; \theta)\}$
and $X^m_t = \{\textbf{x}_{(t, c)}| c \in R^m_t \}$. The loss from ranking with the local context can be formulated as:
\begin{equation}
    \mathcal{L}(\hat{\theta}) = \sum_{t\in\tau} \ell \Bigl ( 
    \bigl \{ y_{(t, c)}, \tilde{f} \bigl (\textbf{x}_{(t, c)} ; \tilde{\theta} \bigr ) \bigr | c \in I(R^m_t, X^m_t, f(\cdot; \theta)) \bigr \}
    \Bigr )  \label{eqn:caltr_emp_loss}
\end{equation}
where $\tilde{f}(\cdot; \tilde{\theta})$ stands for the context-aware scoring function that is parameterised by $\tilde{\theta}$. For the bottom retrieved items, $R^m_t$ is defined as the lowest-$m$ items sorted by the same function $f(\cdot;\theta)$.

In this work, the Transformer-based architecture proposed by \cite{pobrotynContextAwareLearningRank2020} plays the role of the context-aware ranking model $\tilde{f}$. This model is separately trained using the pairwise and listwise losses described in Section \ref{sec:loss_functions}. 
We apply our model to re-rank some top and bottom $m$ subset (which is similar in spirit to \cite{aiLearningDeepListwise2018}), with $m< n$ and where $n=|\mathcal{C}|$. For the remainder of this paper, we drop the subscripts in $\textbf{x}_{(t, c)}$ and $y_{(t, c)}$ for brevity.

\section{Context-Aware Model for Re-ranking}
In this section, we explain the context-aware model as applied to long positions and note that the procedure is similar for shorts. Presented with an initial list of $m$ top retrieved assets produced by a previous ranking algorithm, we want to learn a context-aware scoring function that refines this list by incorporating information across items.

\subsection{Model Inputs}
Inputs comprise of a list of $m$ top retrieved currencies $\textbf{x}$ after an initial round of ranking by some previous LTR algorithm. Each item is first passed through a shared fully connected input layer of size $d_{fc}$.

\subsubsection{Positional Encodings}
To allow the model to leverage item positions provided by the previous ranker, we add positional encodings to the input embeddings. While \cite{vaswaniAttentionAllYou2017} propose both fixed and learnable positional encodings to encode item order in a list, we use the former which is given as:
\begin{align}
    PE_{(p,2i)} & = \sin \Biggl ( \frac{p}{10000^{(2i)/d_{model}}} \Biggr ) \\
    PE_{(p,2i+1)} & = \cos \Biggl ( \frac{p}{10000^{(2i)/d_{model}}} \Biggr )
\end{align}
where $p$ denotes the item's position, $i$ is an element belonging to an embedding dimension index.

\subsection{Encoder Layer}
The Encoder layer facilitates learning the higher-order representations of objects in the list, and it achieves this with the attention mechanism and a feed-forward network alongside various techniques such as dropouts, residual connections and layer normalisation. A schematic of a single encoder layer is displayed on the left of Figure \ref{fig:encoders}.

\subsubsection{Self-Attention Mechanism}
This was introduced in \cite{vaswaniAttentionAllYou2017} and is the key component within the broader model architecture allowing item representations to be constructed. The attention mechanism uses the query, key and value matrices as inputs. Self-attention is a variant of attention where these matrices are identical. In our work, they are representations of the top retrieved items.
We compute attention using the form referred to as Scaled Dot-Product\footnote{There are different ways of computing attention, such as \cite{luong-etal-2015-effective} and \cite{bahdanau2014neural}}:
\begin{equation}
    \textrm{Att}(\textbf{Q},\textbf{K},\textbf{V}) = \textrm{softmax}\Biggl ( \frac{\textbf{QK}^\top}{\sqrt{d_{model}}} \Biggr ) \textbf{V} \label{eqn:att}
\end{equation}
where $\textbf{Q}$ is a query matrix of dimension $d_{model}$ that contains all items in the list, $\textbf{K}$ and  $\textbf{V}$ are respectively the key and value matrices, and lastly $\frac{1}{\sqrt{d_{model}}}$ is a scaling factor to mitigate issues arising from small gradients in the softmax operator.

The model's capacity to learn representations can be further enhanced by ensembling multiple attention modules in what \cite{vaswaniAttentionAllYou2017} refers to as multi-head attention:
\begin{align}
    \textrm{MHA}(\textbf{Q},\textbf{K},\textbf{V}) & = \textrm{concat}(\textrm{head}_1, ... , \textrm{head}_h)\textbf{W}^O \\
    \textrm{head}_i & = \textrm{Att}(\textbf{QW}_i^\textbf{Q}, \textbf{KW}_i^\textbf{K}, \textbf{VW}_i^\textbf{V})
\end{align}
here, each $\textrm{head}_i$ (out of $h$ heads) refers to the $i$th attention mechanism of Equation \ref{eqn:att}, and learned parameter matrices:
\begin{align}
  & \textbf{W}_i^\textbf{Q} \in \mathbb{R}^{{d_{model}}\times d_q}, \quad
  \textbf{W}_i^\textbf{K} \in \mathbb{R}^{{d_{model}}\times d_k}, \nonumber \\
  & \textbf{W}_i^\textbf{V} \in \mathbb{R}^{{d_{model}}\times d_v}, \quad
  \textbf{W}^\textbf{O} \in \mathbb{R}^{hd_v\times d_{model}}  \nonumber
\end{align}
where typically $d_q=d_k=d_v=d_{model}/h$ is used.

\subsubsection{Feed-Forward Network} 
This component within the model introduces non-linearity via its activation and facilitates interactions across different parts of the inputs.

\subsubsection{Stacking Encoder Layers}
A single encoder block $\xi(\cdot)$ can be expressed as:
\begin{align}
    \xi(\textbf{x}) & = \Gamma(\textbf{z} + \delta(\phi(\textbf{z}))) \\
    \textbf{z} & = \Gamma(\textbf{x} + \delta(\textrm{MHA}(\textbf{x})))
\end{align}
where $\Gamma(\cdot)$ and $\delta(\cdot)$ are respectively the layer normalisation and dropout functions. Additionally, $\phi(\cdot)$ represents a projection onto a fully-connected layer, and MHA$(\cdot)$ is the multi-head attention module. Stacking multiple encoder layers iteratively feeds an encoder's output into the next, and this grows the model's ability to learn more complex representations:
\begin{equation}
    \boldsymbol{\xi}(\textbf{x}) = \xi_1 \bigl(...(\xi_N(\textbf{x})) \bigr )\label{eqn:enc_stack}
\end{equation}
where $\boldsymbol{\xi}(\textbf{x})$ in Equation \ref{eqn:enc_stack} denotes a stacked encoder involving a series of $N$ encoder layers.

\subsection{Model Architecture}
\begin{figure}[t!]
  \centering
  \includegraphics[width=\linewidth]{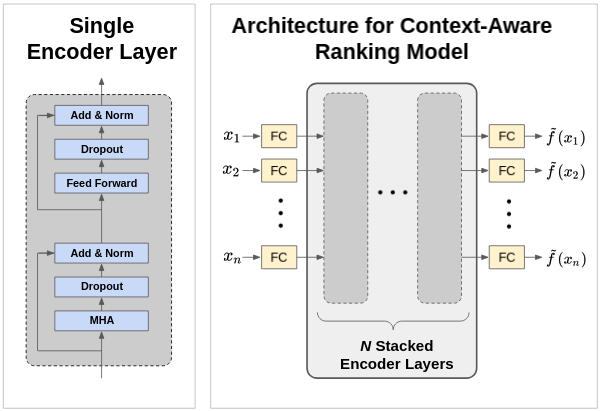}
  \caption{Schematic of a single Encoder Layer (left), and the broader proposed architecture for the context-aware ranking model (right).}
  \label{fig:encoders}
\end{figure}

Following the approach of \cite{pobrotynContextAwareLearningRank2020}, our context-aware ranking model closely resembles the Transformer's encoder with an additional linear projection on the input. Presented with top retrieved currencies provided by an initial ranking algorithm, we pass the input through a fully connected layer of size $d_{fc}$. Next, we feed this through $N$ stacked encoders before sending the results to another fully-connected layer that is used to compute scores.
This entire pipeline can be compactly expressed as:
\begin{equation}
    \tilde{f}(\textbf{x}) = \phi_{\mathrm{output}} \bigl ( \boldsymbol{\xi} (\phi_{\mathrm{input}}(\textbf{x})) \bigr )
\end{equation}
and is graphically represented on the right of Figure \ref{fig:encoders}.

\section{Performance Evaluation}
\subsection{Dataset Overview} \label{sec:pe_data_overview}

We use daily data obtained from the Bank for International Settlements (BIS) \cite{bis_currencies}. From this, daily portfolios are constructed using the same set of 31 currencies in \cite{bazDissectingInvestmentStrategies2015}. Our study spans 2000 to 2020 with the set of currencies all expressed versus the USD. For conditioning our daily data on normal and risk-off market states, we make use of the daily close of VIX historical data from Cboe Global Markets \cite{vix_historical_data}. 

\subsection{Predictor Description} \label{sec:pe_backtest}
We use a simple combination of returns-based features for our predictors:
\begin{enumerate}
    \item \textit{1-month raw returns} -- Based on the best model in \cite{menkhoffCurrencyMomentumStrategies2012}, which involves scoring instruments based on returns calculated over the previous one month.
    \item \textit{Normalised returns} -- Returns over the past 1, 3, 5, 10 and 21-day periods standardised by daily volatility and then scaled to the appropriate time scale.
    \item \textit{MACD-based indicators} -- Final momentum trading indicator along with its constituent raw signals as defined in \cite{bazDissectingInvestmentStrategies2015} and \cite{pohBuildingCrossSectionalSystematic2021}.
\end{enumerate}

\subsection{Loss Functions}\label{sec:loss_functions}

For the loss function $\ell$ used for training the model, we use a pairwise (Pairwise Logistic) loss and a listwise (ListNet) loss. Both losses are commonly used and are described below. In the equations that follow, $\textbf{x}$ refers to a list of inputs, and $f(\textbf{x})$ represents the score vector obtained from the ranking function $f$ and lastly, $\textbf{y}$ denotes the vector of ground-truth labels.

\subsubsection{Pairwise Logistic Loss} Proposed by \cite{arias-nicolasLogisticRegressionbasedPairwise2008}, this loss is defined as:
\begin{equation}
    \ell(\textbf{y}, f(\textbf x)) = \sum^n_{i=1} \sum^n_{j=1}
    \mathbb{I}(y_i > y_j)\log \bigl ( 1 + \exp(f(\textbf{x})_j - f(\textbf{x})_i) \bigr )
\end{equation}
where $\mathbb{I}(\cdot)$ is the indicator function. 
\subsubsection{ListNet Loss} The ListNet loss \cite{caoLearningRankPairwise2007} is a listwise loss and is given as:
\begin{equation}
    \ell(\textbf{y}, f(\textbf{x})) = -\sum_i \textrm{softmax}(\textbf{y})_i \times \log \bigl (\textrm{softmax}(f(\textbf{x}))_i \bigr )
\end{equation}
We next list and discuss both baselines and proposed models used in our study, along with the metrics used to evaluate performance.

\subsection{Models and Comparison Metrics} \label{sec:pe_models}

We first list the benchmark models (with their corresponding shorthand in parentheses) used in this paper. The following 3 models are collectively referred to as conventional benchmarks:
\begin{enumerate}
    \item Random (Rand) -- Buys/sells at random. Included to give some absolute baseline sense of performance.\label{model_list:conventional_first}
    \item Volatility Normalised MACD (Baz) -- Heuristics-based ranker with a sophisticated trend estimator proposed by \cite{bazDissectingInvestmentStrategies2015}.
    \item Multi-Layer Perceptron (MLP) -- Characterises the typical Regress-then-rank model used by contemporary strategies. \label{model_list:conventional_last}
\end{enumerate}
The next 4 models are grouped as LTR benchmarks:
\begin{enumerate}\setcounter{enumi}{3}
    \item Pairwise Logistic Regression (PW) -- Neural network trained using the Pairwise Logistic Loss of  \cite{arias-nicolasLogisticRegressionbasedPairwise2008} \label{model_list:first_ltr}
    \item ListMLE (ML) -- Listwise neural LTR model by \cite{xiaListwiseApproachLearning2008}.
    \item ListNet (LN) -- Listwise neural LTR model by \cite{caoLearningRankPairwise2007}.
    \item LambdaMART (LM) -- Pairwise tree-based ranker by \cite{burgesRankNetLambdaRankLambdaMART2010}. \label{model_list:last_ltr}
\end{enumerate}
With all benchmarks, we follow \cite{bazDissectingInvestmentStrategies2015} --  rebalancing daily and forming equally weighted long/short decile portfolios with six currencies (i.e. using the 3 most extreme pairs on each side).

For the context-aware rankers, we have the following:
\begin{itemize}
  \item Context-aware PW models: PW+P, PW+L -- each is based on the baseline PW model that is re-ranked with the Transformer and calibrated with the pairwise and listwise loss respectively.
  \item Context-aware ML models: ML+P, ML+L.
  \item Context-aware LN models: LN+P, LN+L.
  \item Context-aware LM models: LM+P, LM+L.
\end{itemize}
Finally, we re-rank the top 10 retrieved instruments (i.e., $m=10$) with a context-aware model, and use the resulting top 3 items to construct our long portfolio. For shorts, we re-rank the bottom 10 and select the bottom 3.

The financial and ranking performance of the various models are evaluated over the following metrics:
\begin{itemize}
    \item Profitability: Expected returns ($\mathbb{E}[\textrm{Returns}]$) and hit rate (percentage of positive returns at the portfolio-level obtained over the out-of-sample period).
    \item Risks: Volatility, Maximum Drawdown (MDD) and Downside Deviation.
    \item Financial Performance: Sharpe $\Bigl(\frac{\mathbb{E}[\textrm{Returns}]}{\textrm{Volatility}}\Bigr)$, Sortino $\Bigl(\frac{\mathbb{E}[\textrm{Returns}]}{\textrm{MDD}}\Bigr)$ and Calmar $\Bigl(\frac{\mathbb{E}[\textrm{Returns}]}{\textrm{Downside Deviation}}\Bigr)$ ratios are used as a gauge to measure risk-adjusted performance. We also include the average profit divided by the average loss $\Bigl(\frac{\textrm{Avg. Profits}}{\textrm{Avg. Loss}}\Bigr)$.
    \item Ranking Performance: NDCG@3. This is based on the Normalised Discounted Cumulative Gain (NDCG) \cite{jarvelinIREvaluationMethods2000} which is a graded relevance measure. However, rather than assessing NDCG across the entire cross-section of instruments, we focus on the top/bottom 3 assets which is directly linked to strategy performance.
\end{itemize}

\subsection{Backtest and Training Details}\label{sec:pe_backtest_training}
All LTR models (baseline and context-aware) and the MLP were tuned in blocks of 5-year intervals, with the calibrated weights and hyperparameters fixed and then used for portfolio rebalancing in the following out-of-sample 5-year window. For a given training set, 90\% of the data was used for training and the remainder is reserved for validation. Backpropagation was conducted for a maximum of 100 epochs. For baseline LTR models and the MLP, we used 2 hidden layers with the layer width treated as a tunable hyperparameter. Early stopping was applied to mitigate overfitting, and was triggered when a model's loss on the validation set did not improve for 25 consecutive epochs. Across all models, hyperparameters were tuned using a 50-iterations search with \texttt{HyperOpt} \cite{bergstraHyperoptPythonLibrary2015}. Further details on calibrating the hyperparameters can be found in Section \ref{app:training_details} of the Appendix.  

\subsection{Results and Discussion}

\begin{table*}[t!]\centering
\caption{Performance metrics for all benchmarks models, rescaled to the 15\% annual target volatility. The set of LTR baselines perform just as well if not better than the conventional baselines.}
\label{table:all_benchmarks}
\begin{tabular}{@{}rrrrcrrrr@{}}
\toprule
& \multicolumn{3}{c}{\textbf{Conventional}}& \phantom{abc}& \multicolumn{4}{c}{\textbf{LTR}} \\
\cmidrule{2-4}\cmidrule{6-9} & Rand & Baz & MLP && PW & ML & LN & LM \\
\midrule
E[returns] 	    & -0.010 & 0.086 & \textbf{0.091} && 0.098 & 0.096 & 0.098 & \textbf{**0.129} \\
Volatility 	    &  \textbf{**0.155} & 0.160 & 0.158 && 0.157 & 0.164 & 0.162 & \textbf{0.157} \\
Sharpe Ratio 	& -0.064 & 0.539 & \textbf{0.572} && 0.624 & 0.586 & 0.602 & \textbf{**0.822} \\
Downside Dev.   &  0.108 & \textbf{0.106} & 0.108 && 0.110 & \textbf{**0.106} & 0.108 & 0.110 \\
Max Drawdown    &  0.365 & \textbf{0.365} & 0.543 && \textbf{**0.240} & 0.538 & 0.326 & 0.264 \\
Sortino Ratio 	& -0.092 & 0.814 & \textbf{0.843} && 0.890 & 0.907 & 0.905 & \textbf{**1.179} \\
Calmar Ratio 	& -0.027 & \textbf{0.236} & 0.167 && 0.407 & 0.179 & 0.299 & \textbf{**0.490} \\
Hit-rate 	    &  0.502 & 0.520 & \textbf{0.530} && 0.523 & 0.510 & 0.515 & \textbf{**0.531} \\
AP/AL 		    &  0.980 & \textbf{1.014} & 0.979 && 1.016 & \textbf{**1.068} & 1.047 & 1.018 \\
\bottomrule\end{tabular}
\end{table*}

\begin{table*}[t!]\centering
\caption{Performance metrics for all LTR models (original and context-aware), rescaled to the 15\% annual target volatility. Context-aware LTR models outperform across all measures.}
\label{table:all_ltr_models}
\begin{tabular}{@{}rrrrrrrrrrrrrrrr@{}}
\toprule
& \multicolumn{3}{c}{\textbf{PW-based models}}& \phantom{abc} & \multicolumn{3}{c}{\textbf{ML-based models}}& \phantom{abc} &\multicolumn{3}{c}{\textbf{LN-based models}} & \phantom{abc} & \multicolumn{3}{c}{\textbf{LM-based models}} \\
\cmidrule{2-4}\cmidrule{6-8}\cmidrule{10-12}\cmidrule{14-16} 
& PW & PW+P & PW+L && ML & ML+P & ML+L && LN & LN+P & LN+L && LM & LM+P & LM+L \\
\midrule
E[returns]      & 0.098 & \textbf{0.141} & 0.117 &&  0.096 &  0.116 &  \textbf{0.120} &&  0.098 &  \textbf{0.157} &  0.133 && 0.129 & 0.148 & \textbf{**0.157} \\
Volatility      & 0.157 & \textbf{0.156} & 0.157 &&  0.164 &  0.158 &  \textbf{0.157} &&  \textbf{0.162} &  0.165 &  0.165 && 0.157 & 0.157 & \textbf{**0.156} \\
Sharpe Ratio    & 0.624 & \textbf{0.905} & 0.746 &&  0.586 &  0.729 &  \textbf{0.764} &&  0.602 &  \textbf{0.948} &  0.806 && 0.822 & 0.941 & \textbf{**1.008} \\
Downside Dev.   & 0.110 & \textbf{0.107} & 0.108 &&  \textbf{0.106} &  0.110 &  0.107 &&  \textbf{0.108} &  0.109 &  0.110 && 0.110 & 0.109 & \textbf{**0.106} \\
Max Drawdown    & 0.240 & \textbf{**0.234} & 0.287 &&  0.538 &  \textbf{0.533} &  0.540 &&  0.326 &  0.285 &  \textbf{0.239} && 0.264 & \textbf{0.236} & 0.279 \\
Sortino Ratio   & 0.890 & \textbf{1.315} & 1.086 &&  0.907 &  1.050 &  \textbf{1.119} &&  0.905 &  \textbf{1.431} &  1.210 && 1.179 & 1.357 & \textbf{**1.479} \\
Calmar Ratio    & 0.407 & \textbf{0.604} & 0.410 &&  0.179 &  0.217 &  0\textbf{.223} &&  0.299 &  0.549 &  \textbf{0.559} && 0.490 & \textbf{**0.626} & 0.564 \\
Hit-rate        & 0.523 & \textbf{0.532} & 0.518 &&  0.510 &  0.519 &  \textbf{0.523} &&  0.515 &  \textbf{0.531} &  0.514 && 0.531 & \textbf{**0.535} & 0.535 \\
AP/AL           & 1.016 & 1.025 & \textbf{1.056} &&  \textbf{1.068} &  1.050 &  1.038 &&  1.047 &  1.047 &  \textbf{**1.093} && 1.018 & 1.022 & \textbf{1.032} \\
\bottomrule\end{tabular}
\end{table*}

\begin{figure*}[t!]
  \centering
  \includegraphics[width=\linewidth]{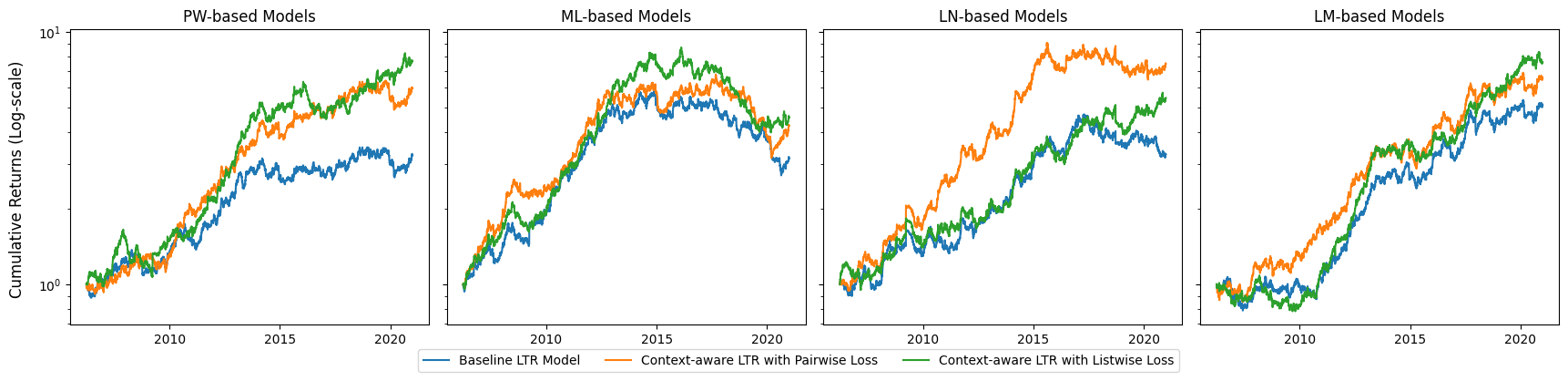}
  \caption{Cumulative returns for both baseline LTR and context-aware LTR models, rescaled to target volatility. Models are grouped by their respective baseline ranking models. For example, PW and both context-aware PW models re-ranked with the pairwise and listwise losses (PW+P and PW+L respectively) are grouped together. The plots collectively indicate that incorporating context-awareness improves the cumulative wealth of all original LTR strategies.}\label{fig:perf}
\end{figure*}

\begin{table*}[t!]\centering
\caption{Long/Short ranking performance for conventional and LTR baselines. Focusing on the average NDCG@3 across all rebalances, the LTR models are generally as accurate if not more accurate than the conventional benchmarks.}
\label{table:ranking_metrics_benchmarks}
\begin{tabular}{@{}rrrrcrrrr@{}}
\toprule
& \multicolumn{3}{c}{\textbf{Conventional}}& \phantom{abc}& \multicolumn{4}{c}{\textbf{LTR}} \\
\cmidrule{2-4}\cmidrule{6-9} & Rand & Baz & MLP && PW & ML & LN & LM \\
\midrule
NDCG@3 	    & 0.553 & 0.556 & \textbf{0.556} && 0.559 & 0.556 & 0.555 & \textbf{**0.560} \\
\bottomrule\end{tabular}
\end{table*}

\begin{table*}[t!]\centering
\caption{Long/Short ranking performance for baseline LTR and context-aware LTR models conditioning on market environments. Models are grouped by their respective baseline ranking models. Across both normal and risk-off market states, the context-aware ranking models are able to more accurately select the winners and losers, as measured by the NDCG@3 averaged over each state. This difference in ability is greater in risk-off states.}
\label{table:ranking_metrics_NDCG3}
\begin{tabular}{@{}rrrrrrrrrrrrrrrr@{}}
\toprule
& \multicolumn{3}{c}{\textbf{PW-based models}}& \phantom{abc} & \multicolumn{3}{c}{\textbf{ML-based models}}& \phantom{abc} &\multicolumn{3}{c}{\textbf{LN-based models}} & \phantom{abc} & \multicolumn{3}{c}{\textbf{LM-based models}} \\
\cmidrule{2-4}\cmidrule{6-8}\cmidrule{10-12}\cmidrule{14-16} 
& PW & PW+P & PW+L && ML & ML+P & ML+L && LN & LN+P & LN+L && LM & LM+P & LM+L \\
\midrule
Normal     & 0.558 & \textbf{0.560} & 0.558 && 0.556 & 0.557 & \textbf{0.558} && 0.554 & \textbf{0.557} & 0.555 && 0.560 & \textbf{**0.560} & 0.560 \\
Risk-Off    & 0.564 & 0.572 & \textbf{0.574} && 0.549 & \textbf{0.566} & 0.555 && 0.560 & 0.569 & \textbf{**0.580} && 0.550 & \textbf{0.556} & 0.551 \\
\bottomrule\end{tabular}
\end{table*}

\begin{table*}[t!]\centering
\caption{Sharpe ratios for LTR and context-aware LTR models over normal and risk-off market states. Context-aware ranking models typically possess higher Sharpe ratios over both market environments.}
\label{table:risk_onoff_sharpe_ratios}
\begin{tabular}{@{}rrrrrrrrrrrrrrrr@{}}
\toprule
& \multicolumn{3}{c}{\textbf{PW-based models}}& \phantom{abc} & \multicolumn{3}{c}{\textbf{ML-based models}}& \phantom{abc} &\multicolumn{3}{c}{\textbf{LN-based models}} & \phantom{abc} & \multicolumn{3}{c}{\textbf{LM-based models}} \\
\cmidrule{2-4}\cmidrule{6-8}\cmidrule{10-12}\cmidrule{14-16} 
& PW & PW+P & PW+L && ML & ML+P & ML+L && LN & LN+P & LN+L && LM & LM+P & LM+L \\
\midrule
Normal    & 0.705 & \textbf{0.872} & 0.724 && 0.691 & 0.775 & \textbf{0.870} && 0.573 & \textbf{0.936} & 0.735 && 0.913 & 1.041 & \textbf{**1.065} \\
Risk-off    & -0.652  & \textbf{1.495} & 1.137 && -1.174 & \textbf{0.109} & -1.049 && 1.151 & 1.200 & \textbf{**2.126} && -0.706 & -0.720 & \textbf{0.003} \\
\bottomrule\end{tabular}
\end{table*}

Table \ref{table:all_benchmarks} consolidates performance figures for both conventional and LTR benchmarks, while Table \ref{table:all_ltr_models} compares the performance of the same LTR models against their context-aware counterparts that are produced by re-ranking under the pairwise and listwise losses described in Section \ref{sec:loss_functions}\footnote{For all tables in this section, bold figures represents the best statistic for models contained within the same group for its respective performance measure. Doubly asterisked figures are the best statistics across \textit{all} models for its respective performance measure.}.
For the results contained in both tables, we perform an additional layer of volatility scaling at the portfolio level -- facilitating comparison and aligning returns with our 15\% volatility target. All returns are computed in the absence of transaction costs to focus on the models' raw predictive ability.

With the set of baselines in Table \ref{table:all_benchmarks}, we see that the group of models employing LTR perform just as well if not better than the other benchmarks over various measures of performance. Among the set of LTR benchmarks, LambdaMART dominates all other rankers across most measures of performance -- echoing \cite{qinAreNeuralRankers2021} who document its superiority over neural rankers on three publicly used LTR data sets. We also note the MLP marginally outperforming Baz despite its relatively sophisticated construction involving neural networks -- likely a consequence of over-fitting and the sub-optimality of the regress-then-rank approach \cite{pohBuildingCrossSectionalSystematic2021}.

From the results in Table \ref{table:all_ltr_models}, we observe that the set of context-aware models possess Sharpe ratios that are larger by approximately 30\% on average; On other risk and performance-based measures, they are generally comparable if not better. In Figure \ref{fig:perf}, it is clear that these models also have higher cumulative returns. Taken together, it is clear that our proposed set of context-aware rankers are superior to the benchmarks.

\subsubsection{Ranking and Strategy Performances over Different Market States}

To gauge ranking accuracy across models, we use the Normalised Discounted Cumulative Gain (NDCG) \cite{jarvelinIREvaluationMethods2000} which is a measure of graded relevance and is commonly used in the IR literature. In particular, we focus on the aggregated NDCG@3 computed over the top/bottom 3 currency pairs which assesses the models' accuracy in forming the traded portfolios. From Table \ref{table:ranking_metrics_benchmarks} which averages this metric across for each benchmark over all rebalancing periods, we see that the (globally learned) LTR baselines typically possess values that are comparable to if not higher than the conventional models -- empirically supporting the point that CSM strategies developed using LTR can sort more accurately on average.

To investigate the extent that context-awareness is able to improve ranking accuracy over different trading conditions, we condition strategy performance on (i) normal and (ii) risk-off market states. We define a risk-off state similar to \cite{debockBehaviorCurrenciesRiskoff2015} which is an instance when the VIX is 5\%\footnote{\cite{debockBehaviorCurrenciesRiskoff2015} use 10\% as their threshold.} points higher than its rolling 60-day moving average; Normal states are then simply states which are not risk-off states. With this definition, approximately 4\% of our data is categorised as risk-off states. 
Table \ref{table:ranking_metrics_NDCG3} indicates that the context-aware ranking models are able to more precisely select instruments over both normal and risk-off environments, with larger differences observed for the latter (e.g. between LN and LN+L). This superiority is mirrored in Table \ref{table:risk_onoff_sharpe_ratios}: the context-aware strategies trade at higher Sharpe ratios, with the largest improvements coinciding with the widest disparities exhibited in Table \ref{table:ranking_metrics_NDCG3}. 
Importantly, these statistics establish that incorporating context-awareness into LTR models results in has a pronounced effect on ranking accuracy -- not just on average, but over normal and risk-off market states. This in turn leads to strategies with superior Sharpe ratios.

\section{Conclusion}

Cross-sectional momentum (CSM) strategies developed using Learning to Rank (LTR) algorithms are able to outperform contemporary methods for currency momentum by producing asset rankings that are more accurate on average. However, the underlying ranking algorithms at the heart of these models are typically globally learned and ignore the differences between feature distributions over different periods when the portfolio is rebalanced. This flaw renders the broader strategy susceptible to producing inaccurate rankings, potentially at critical points in time when accuracy is actually needed the most. For example, this can happen during risk-off episodes -- leading to large, unwanted portfolio drawdowns.
We rectify this shortcoming with a context-aware LTR model that is based on the Transformer architecture. The model encodes features of the top/bottom retrieved assets provided by a previous ranking algorithm, learns the local ranking context, and uses this to refine the original list.

Backtesting on a slate of 31 currencies, our proposed methodology increases the Sharpe ratio by around 30\% and significantly enhances various performance metrics. When we condition on trading environment, our approach similarly improves the Sharpe ratio in both normal and risk-off market states.
New directions for future work includes conducting a comprehensive performance study of context-aware LTR models across different data sets (e.g. different asset classes, higher frequency LOB data), as well as further innovating on the underlying Transformer-based architecture for (re-)ranking.

\begin{acks}
We would like to thank the Oxford-Man Institute of Quantitative Finance for financial and computing support and SR thanks the UK Royal Academy of Engineering.
\end{acks}

%
\bibliographystyle{ACM-Reference-Format}
\bibliography{refs}

\appendix

\section{Currency Dataset Details}
Following \cite{bazDissectingInvestmentStrategies2015}, we work with 31 currencies all expressed versus the USD over the period 2-May-2000 to 31-Dec-2020. The full currency list is as follows:

\begin{itemize}
    \item G10: AUD, CAD, CHF, EUR, GBP, JPY, NOK, NZD, SEK, USD
    \item EM Asia: HKD, INR, IDR, KRW, MYR, PHP, SGD, TWD, THB
    \item EM Latam: BRL, CLP, COP, PEN, MXN
    \item CEEMEA: CZK, HUF, ILS, PLN, RUB, TRY
    \item Africa: ZAR
\end{itemize}

In order to reduce the impact of outliers, we winsorise the data by capping and flooring it to be within 5 times its exponentially weighted moving (EWM) standard deviation from its EWM average that is calculated using a 252-day span.

\section{Additional Training Details}\label{app:training_details}

\textit{Python Libraries:} LambdaMART uses {\tt XGBoost} \cite{chenXGBoostScalableTree2016}. The MLP, Pairwise Regression Model, ListMLE, ListNet as well as the Transformer underlying all context-aware ranking models (e.g. PW+P, PW+L) are developed using {\tt Tensorflow} \cite{abadiTensorFlowLargescaleMachine2015}. 

\textit{Hyperparameter Optimisation:} Hyperparameters are tuned using {\tt HyperOpt} \cite{bergstraHyperoptPythonLibrary2015}. For LambdaMART, we refer to the hyperparameters as they are named in the {\tt XGBoost} library.

\vspace{8pt}
\textbf{Multi-layer Perceptron (MLP):}
\begin{itemize}
    \item Dropout Rate -- [0.0, 0.2, 0.4, 0.6, 0.8]
    \item Hidden Width -- [16, 32, 64, 128, 256]
    \item Learning Rate -- $[10^{-7}, 10^{-6}, 10^{-5}, 10^{-4}, 10^{-3}]$
\end{itemize}

\vspace{8pt}
\textbf{ListNet:}
\begin{itemize}
    \item Dropout Rate -- [0.0, 0.2, 0.4, 0.6, 0.8]
    \item Hidden Width -- [8, 16, 32, 64, 128]
    \item Learning Rate -- $[10^{-5}, 10^{-4}, 10^{-3}, 10^{-2}, 10^{-1}]$
\end{itemize}

\vspace{8pt}
\textbf{ListMLE, Pairwise Logistic Regression}
\begin{itemize}
    \item Dropout Rate -- [0.0, 0.2, 0.4, 0.6, 0.8]
    \item Hidden Width -- [8, 16, 32, 64, 128]
    \item Learning Rate -- $[10^{-7}, 10^{-6}, 10^{-5}, 10^{-4}, 10^{-3}]$
\end{itemize}

\vspace{8pt}
\textbf{LambdaMART:}
\begin{itemize}
    \item `objective' -- `rank:pairwise'
    \item `eval\_metric' -- `ndcg'
    \item `eta' -- $[10^{-5}, 10^{-4}, 10^{-3}, 10^{-2}, 10^{-1}]$
    \item `num\_boost\_round' -- $[5, 10, 20, 40, 80, 160, 320]$
    \item `max\_depth' -- $[2, 4, 6, 8, 10]$
    \item `reg\_alpha' -- $[10^{-5}, 10^{-4}, 10^{-3}, 10^{-2}, 10^{-1}]$
    \item `reg\_lambda' -- $[10^{-5}, 10^{-4}, 10^{-3}, 10^{-2}, 10^{-1}]$
    \item `tree\_method' -- `gpu\_hist'
\end{itemize}

\vspace{8pt}
\textbf{Transformer for Context-aware LTR}
\begin{itemize}
    \item $d_{fc}$ -- [16, 32, 64, 128, 256]
    \item $d_{ff}$ -- [16, 32, 64, 128, 256]
    \item Dropout Rate -- [0.0, 0.2, 0.4, 0.6, 0.8]
    \item Learning Rate -- $[10^{-5}, 10^{-4}, 10^{-3}, 10^{-2}, 10^{-1}]$
    \item No. of Encoder Layers -- [1, 2, 3, 4]
    \item No. of Attention Heads -- 1
\end{itemize}

\end{document}